\documentclass[twocolumn,showpacs,preprintnumbers,amsmath,amssymb]{revtex4}

\usepackage{graphicx}
\usepackage{dcolumn}
\usepackage{bm}

\begin{document}

\title{Theory of plasmon decay in dense plasmas and warm dense matter}

\author{S. Son}
\affiliation{18 Caleb Lane, Princeton, NJ 08540}
\author{S. Ku\footnote[1]{sku@cims.nyu.edu}}
\affiliation{Courant Institute of Mathematical Sciences, New York University, New York, NY 10012}
\author{Sung Joon Moon\footnote[2]{Current Address: 28 Benjamin Rush Ln. Princeton, NJ 08540}}
\affiliation{PACM, Princeton University, Princeton, NJ 08544}
\date{\today}

\begin{abstract}
The prevalent Landau damping theory for  classical plasmas
does not fully explain the Langmuir wave decay in dense plasmas.
A dielectric function theory adapted from the condensed matter physics is extended
in order to be applied to dense plasmas and warm dense matter.
This theory, accounting for the Umklapp process, predicts much higher decay
rates than the Landau damping theory, which is in better agreement with the
existing experimental data obtained from the metals.
It is demonstrated that
this strong plasmon decay leads to the existence of a parameter regime where the
backward Raman scattering is unstable while the forward Raman scattering is stable.  
By means of the backward Raman compression, intense x-ray pulses can be created
in this regime.
\end{abstract}
\pacs{52.35.-g, 71.45.Gm, 42.55.Vc, 78.70.Ck} 

\maketitle

\section{Introduction}

Understanding the warm dense matter and dense plasmas has recently become
more important in the context of various practical problems, including the
inertial confinement fusion~\cite{lindl, Tabak}.
In these media, many interesting phenomena observed in  classical plasmas
would still occur, but they get modified due to the quantum diffraction and the
degeneracy~\cite{sonpla, sonprl, Kue, sonlandau}.
We consider one such phenomenon, the decay of the Langmuir wave, which plays an
important role in various applications~\cite{surface, photo, nano, wake, Fisch, Fisch2}.  

The Landau damping theory is often used to predict the plasmon decay in classical plasmas.
On the other hand, the plasmon decay rate in metals is computed by complicated
quantum theories accounting for the local-field correction or the density functional
approach~\cite{Ichimaru,DuBois}. 
However, neither of them is sufficient to predict the plasmon decay in warm dense
matter or dense plasmas.
In this paper, we introduce a theoretical framework to calculate the plasmon decay
in dense plasmas by adapting the dielectric function approach used in the condensed matter physics.
It is suggested that the modification of the free-electron wave packet due to
the presence of ions is considerable, which in turn modifies the plasmon decay profiles.  
It is shown that the decay rate of the long wavelength plasmon is much higher than the
prediction by the Landau damping theory, and that the rate is finite and non-negligible
in the consideration of the Raman compression.  
It leads us to show that there exists a parameter regime where the backward Raman
scattering (BRS) is unstable while the forward Raman scattering (FRS) is stable,
enabling the x-ray compression without a premature pump depletion.

The rest of the paper is organized as follows.
The basic classical and quantum-mechanical Landau damping theory is summarized
in Sec. II, and why the prevalent Landau damping theory is insufficient for metals
and dense plasmas is argued in Sec. III.
A dielectric function formalism, adapted from the condensed matter physics research
is introduced and extended to be applicable to the dense plasmas and warm dense
matter (Secs. IV and V) 
The plasmon damping rate is computed in this new framework and the existence of
a regime in dense plasmas where the BRS can be unstable while the FRS is stable
is demonstrated (Sec. VI), and then it is concluded in Sec. VII.

\section{Landau Damping Theory}

Consider a wave of the form $\phi(\mathbf{x},t) = \phi \exp(i\mathbf{k}\cdot \mathbf{x} -i\omega t) + \phi^* \exp(-i\mathbf{k}\cdot \mathbf{x} +i\omega t )$.
The classical analysis of the Landau damping rate, $\gamma_{\mathrm{cl}}$,
for small $e\phi$ is~\cite{Landau} 
\begin{equation}
 \frac{\gamma_{\mathrm{cl}}}{\omega} = \frac{\pi}{2} \frac{\omega_{\mathrm{pe}}^2}{k^2} \frac{\partial f(\omega/k)}{\partial v} \mathrm{,} \label{eq:classical}
\end{equation} 
where $\mathbf{k}$ ($\omega$) is the wave vector (frequency),
$\omega_{\mathrm{pe}} = \sqrt{4\pi n_e e^2/m_e}$ is the plasma frequency,
$n_e$ ($m_e$) is the electron density (mass),
and $f$ is the electron probability distribution function satisfying
$\int f d^3 \mathbf{v} = 1 $.
The damping rate $\gamma$ can be alternatively derived in the dielectric function
formalism,
\begin{equation}
 \frac{\gamma}{\omega}  =\frac{1}{2} \mathrm{Im}\left[\epsilon(\mathbf{k},\omega)\right] \mathrm{,}\label{eq:die}
\end{equation}
where $\epsilon$ is the well-known dielectric function obtained through the random phase approximation;
\begin{equation}
 \epsilon(k,\omega) = 1 + \frac{4 \pi n_e e^2 }{m_e k^2} \int \frac{\mathbf{k} \cdot \nabla_{v} f }{ \omega- \mathbf{k}\cdot \mathbf{v}} \mathrm{.} \label{eq:cla2}
\end{equation}
In the regime where the x-ray compression would be feasible through the BRS,
$n_e$ is so high that the electron quantum diffraction cannot be neglected.
In such a case, the quantum version of the dielectric function, the so-called Lindhard
function~\cite{Lindhard}, needs to be used;
\begin{equation} 
\epsilon(\mathbf{k},\omega) = 1 + \frac{\omega_{\mathrm{pe}}^2}{k^2} \frac{m_e}{\hbar} \int 
\frac{ f(E_f) - f(E_i)}{\omega - (E_f - E_i)/\hbar } d^3 \mathbf{v}\mathrm{,} \label{eq:quantum}
\end{equation} 
where $E_i = m_e \mathbf{v}_i^2/2$ ($E_f  =  m_e \mathbf{v}_f^2/2$) is the electron kinetic energy of the momentum $m_e \mathbf{v}_i = m_e \mathbf{v} - \hbar \mathbf{k}/2$ ($m_e \mathbf{v}_f =m_e \mathbf{v} + \hbar \mathbf{k}/2 $). 
In order to derive Eq.~(\ref{eq:quantum}), we first solve the time-dependent Schroedinger
equation using a perturbation analysis for each free electron wave packet and arrive at 
\begin{eqnarray}
 |\mathbf{q}\rangle  =   |\mathbf{q}\rangle_0  +  \frac{ e\phi \exp(i\mathbf{k}\cdot \mathbf{x} -i\omega t)}{ \hbar \omega - E(\mathbf{q}+\mathbf{k}) + E(\mathbf{q}) } |\mathbf{q}+\mathbf{k}\rangle_0 \nonumber \\ \nonumber \\ 
  -  \frac{ e\phi^* \exp(-i\mathbf{k}\cdot \mathbf{x} + i\omega t) }{ \hbar \omega - E(\mathbf{q}-\mathbf{k}) + E(\mathbf{q}) } |\mathbf{q}-\mathbf{k} \rangle_0 \nonumber 
 \mathrm{,}  
\end{eqnarray}
where  $E(\mathbf{p}) = \hbar^2 |\mathbf{p}|^2/2m_e$, $|\ \rangle_0$ denotes
the free electron wave function, and $| \ \rangle$ denotes the perturbed wave function.
The perturbed electron density is given as 
\begin{equation}
\delta n_e(\mathbf{x}, t) =  \int \frac{d^3 \mathbf{q} }{(2\pi)^3} \left[g(\mathbf{q}) |\langle \mathbf{x}|\mathbf{q}\rangle|^2 \right] \mathrm{,}\nonumber 
\end{equation}
where $g(\mathbf{q}) $, satisfying $\int   g(\mathbf{q}) d^3\mathbf{q} / (2\pi)^3 = n_e$,
is the occupation number.
$\delta n_e(\mathbf{k}, \omega)$ reads 
\begin{equation}
\delta n_e(\mathbf{k}, \omega) 
 = e\phi \int \frac{d^3\mathbf{q}}{ (2\pi)^3} g(\mathbf{q})  \frac{f(\mathbf{q}+\mathbf{k}) - f(\mathbf{q})}{  \hbar \omega - E(\mathbf{q}+\mathbf{k}) + E(\mathbf{q}) } \mathrm{.} \label{eq:perturb} 
\end{equation}
The dielectric function, to the first order in $\phi$, is
\begin{equation}
 \epsilon(\mathbf{k},\omega)= 1 + \frac{4 \pi e^2}{k^2} \alpha(\mathbf{k},\omega) \nonumber \mathrm{,}
\end{equation}
where $\alpha(\mathbf{k},\omega) =  \delta n_e(\mathbf{k},\omega)/e\phi$ is the susceptibility.
The imaginary part of Eq.~(\ref{eq:quantum}) is
\begin{equation} 
\mathrm{Im}[\epsilon] =\frac{\omega_{\mathrm{pe}}^2}{k^2} \frac{m_e}{\hbar} \int \pi \delta( \omega - \frac{E_f - E_i}{\hbar} )
 \left(f(E_f) - f(E_i)\right) d^3 \mathbf{v}\mathrm{.} \nonumber 
\end{equation}  
Note that the integration is done in the velocity space which is simply related to the
momentum space by $ \hbar \mathbf{q} / m_e =\mathbf{ v} $. 
In the limit $\hbar \cong 0$,
the above equation is reduced to Eq.~(\ref{eq:classical}).

\section{Langmuir Wave Damping in Metals and Breakdown of the Landau Damping theory} 

\begin{figure}
\scalebox{1.2}{
\includegraphics[width=0.795\columnwidth]{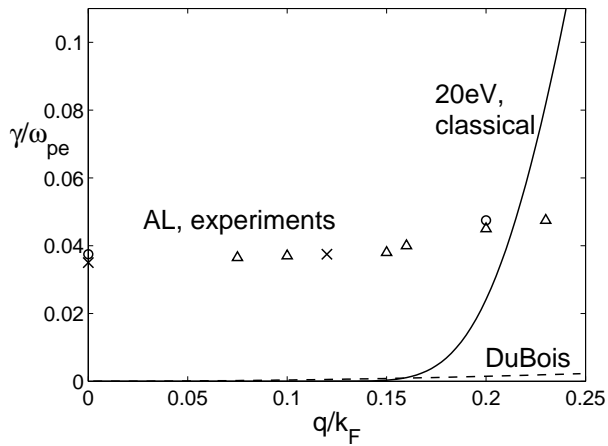}}
\caption{\label{fig:1}
Comparison of various damping rates of a plasma of $n_e = 1.8\times 10^{23} \mathrm{cm^{-3}}$,
which is a reproduction of Fig.~1 in Ref.~\cite{sonpre}.
The prediction from the classical dielectric function (solid line) for $T_e = 20 \ \mathrm{eV} $
is compared with that of the DuBois' theory (dashed line) for $T_e=0$.
The experimental measurements for AL are adapted from Fig.~3 in
Ref.~\cite{Sturm2}, which are originally from Ref.~\cite{Zacharias}
(crosses), Ref.~\cite{Urner-Wille} (circles), and Ref.~\cite{Kloos}
(triangles).
}
\end{figure}

The Landau damping rate in the free-electron plasma can be computed by either the classical
dielectric function (Eq.~(\ref{eq:classical})) or the degenerate Lindhard dielectric
function (Eq.~(\ref{eq:quantum})).
The rate in rare dense plasmas is accurately predicted by these approaches,
However, the rate in dense plasmas, warm dense matter, or metals is not.
For an instance, the comparison between experimentally obtained damping rates
in Al and theoretical computations exhibit large deviation (Fig.~\ref{fig:1}).
The prediction by the Lindhard dielectric function~\cite{Lindhard, Nagy},
Eq.~(\ref{eq:quantum}), is essentially zero in this range of $k$, and is not
visible in the figure. 
The damping rate from the DuBois' theory~\cite{DuBois}, accounting for the dynamical
correlations, predicts the damping to be proportional to $(k/k_F)^2$ in the long
wavelength limit where $k$ goes to zero.
However, the data from the electron stopping experiments in
metals~\cite{Gibbons,Thira} suggest that the rate is finite in this limit,
which is inconsistent with the prediction of the aforementioned
theories~\cite{Sturm, Ku, Ku2, Pitarke}.  
The experimental decay rate for $k < 0.5 k_F$ can be estimated as 
\begin{equation}
\nu_3(k) = \eta(k) \omega_{\mathrm{pe}} / \sec \label{eq:lang} \mathrm{.}
\end{equation}
where $\eta(k) = \eta_0 + d\eta/dk^2 (k/k_F)^2$.
For a typical metal, $ 0.02 < \eta_0 < 0.2$ and $ d\eta/dk \cong a \eta_0 $,
where $ 2 < a < 10$~\cite{Gibbons}.  
It is tempted to explain that the finite plasmon decay rate for $k \cong 0$ 
arises from the electron-ion collisions.
However, it is difficult to estimate the decay rate by this argument.
Furthermore, the electron-ion collisions get diminished in degenerate plasmas by
the electron diffraction and the degeneracy~\cite{sonpla, sonprl}, and it is not
clear how the quantum effect can be incorporated into the plasmon damping.

The electron wave packets in metals are distorted from those of the free electrons
through the strong interaction between the electrons and the ion lattices.
Adler~\cite{adler} computed the electronic dielectric function in the presence of
the metal ion lattices,
by expanding the wave packet in terms of the reciprocal lattice wave vector
and using the quantum random phase approximation.
Sturm~\cite{Sturm} and Hasegawa~\cite{Hasegawa} applied this dielectric function formalism
to Eq.~(\ref{eq:die}) and obtained the Langmuir wave decay rate, which is in
good agreement with the experimental data~\cite{Gibbons}.
In particular, in the long wavelength limit, the damping rate is predicted to
be finite. 
The dominant mechanism of the decay in Alkali metals is shown to be the Umklapp
process; an electron inside the Fermi shell gets excited by the Langmuir wave
outside the shell, obtaining the momentum of $\mathbf{k}+\mathbf{G}$, where $\mathbf{k}$ is the Langmuir
wave vector and $\mathbf{G}$ is the reciprocal vector. 

In dense plasmas or warm dense matter, the electron density is so high that many physical
properties are similar to those in metals, and so is the decay profile of the plasmon
damping;
the ion density is also high, and the distortion of the electron wave packet and the
Umklapp excitation become significant.  

\section{Dielectric Function in Metals and its Extension to Dense Plasmas}

In this Section, the dielectric function theory developed for the metals is briefly
reviewed and extended for the case of dense plasmas and warm dense matter. 
In both cases, the electron eigenstates are assumed to deviate strongly
from the free-electron wave packets, $|\sigma \rangle$, where $\sigma$ is an index.
In the presence of the potential of the form
$\phi(\mathbf{x},t) = \phi \exp(i\mathbf{k}\cdot \mathbf{x} -i\omega t) + \phi^* \exp(-i\mathbf{k}\cdot \mathbf{x} +i\omega t )$, the wave packet is modified to be
\begin{eqnarray}
 \parallel\sigma\rangle = |\sigma \rangle  + \sum_{\sigma_1}\frac{ e\phi}{ \hbar \omega - E_{\sigma_1} + E_{\sigma}} |\sigma_1 \rangle  \langle \sigma_1| \exp(i\mathbf{k}\cdot \mathbf{x})| \sigma\rangle  \nonumber \\
+ \sum_{\sigma_1}\frac{ e\phi^*}{ -\hbar \omega + E_{\sigma_1} - E_{\sigma}} 
|\sigma_1 \rangle \langle \sigma_1| \exp(-i\mathbf{k}\cdot \mathbf{x}) |\sigma\rangle  \nonumber \mathrm{,} \\ \label{eq:sigma} \nonumber  
\end{eqnarray}
where the perturbation is assumed to be weak and the perturbation theory of the first order is used.
$|~\rangle$ denotes the original eigenstate and $\parallel~\rangle $ denotes the perturbed eigenstate.
The perturbation in the density obtained from the above relation is
\begin{eqnarray}
\delta n(k,\omega) = \left[ \sum_{\sigma_1} f(\sigma_1) |\langle\sigma_1 \parallel \exp(i\mathbf{k}\cdot\mathbf{x}) \parallel\sigma_1 \rangle|^2\right] \nonumber \\ 
= \sum_{\sigma_1, \sigma_2} e\phi  (f(\sigma_1) -f(\sigma_2)) \beta(\sigma_1, \sigma_2, \mathbf{k}, \omega ) \mathrm{,} \nonumber \\  \nonumber 
\end{eqnarray}
where 
\begin{equation}
  \beta(\sigma_1, \sigma_2) =
\frac{\langle\sigma_1|\exp(-i\mathbf{k}\cdot \mathbf{x})|\sigma_2\rangle\langle\sigma_2|\exp(i\mathbf{k}\cdot \mathbf{x})|\sigma_1\rangle}{ \hbar \omega - E_{\sigma_1} + E_{\sigma_2}}  \mathrm{.} \nonumber 
\end{equation}
The dielectric function, up to the first order in $\phi$, is given to be 
\begin{eqnarray}
 \epsilon(k,\omega) = 1 + \frac{4\pi n_e e^2}{k^2} \sum_{\sigma_1, \sigma_2} e\phi  (f(\sigma_1) -f(\sigma_2)) \nonumber \\ \times  
\frac{\langle\sigma_1|\exp(-i\mathbf{k}\cdot \mathbf{x})|\sigma_2\rangle\langle\sigma_2|\exp(i\mathbf{k}\cdot \mathbf{x})|\sigma_1\rangle}{ \hbar \omega - E_{\sigma_1} + E_{\sigma_2}} \mathrm{.} \label{eq:adler}
\end{eqnarray}
With an appropriate choice for the eigenstate in a given condition, Eq.~(\ref{eq:adler})
can be used in various situations.
For example, Sturm~\cite{Sturm,Sturm2} used the following eigenstate
\begin{equation} 
|\sigma\rangle = |\mathbf{q}\rangle + \sum_{\mathbf{q}_1 \neq \mathbf{q}} |\mathbf{q}_1 \rangle \frac{\langle \mathbf{q}_1 | V |\mathbf{q}\rangle}{ E_{q} - E_{q_1}} \mathrm{,} \nonumber  
\end{equation}
where $\mathbf{q}$ and $\mathbf{q}_1$ are the wave vectors,
and $\langle \mathbf{q}_1 | V |\mathbf{q}\rangle$ is the pseudo-potential
which is experimentally measurable.
It should be noted that, in case of metals, $\langle \mathbf{q}_1 | V |\mathbf{q}\rangle$
becomes finite only when $\mathbf{q}_1 - \mathbf{q}$ is a reciprocal vector. 

Now we apply the above formalism to dense plasmas or warm dense matter,
where unlike metals, the ions can move as individual particles.
Since the ions move much slower than the electrons and the Langmuir wave frequency
is much faster than the ion relaxation time, we assume that the ions are spatially
frozen at $\mathbf{X}_i$.
We first consider the dynamical property of the electrons in the presence of the
frozen ions, and then obtain the average dynamics by averaging over the probability
distribution of $\mathbf{X}_i$, following the Born-Oppenheimer's approximation.
Without loss of generality, $\mathbf{X}_i$ can be assumed to have the following
correlation average:
\begin{equation}
\langle \sum_{i,j}\frac{\exp(i\mathbf{s}\cdot (\mathbf{X}_i-\mathbf{X}_j))}{V} \rangle= n_I(\mathbf{s})  \mathrm{,} \nonumber 
\end{equation}
where $V$ is the volume of the region under consideration, and $n_I(\mathbf{s})$ is the static
two-point correlation function of the ions. For independent ions, $n_I(\mathbf{s}) = n_I$.
Only one species of ions of the charge $Z$ is considered.
When the ion positions are fixed, the electron's free wave eigenfunction is modified to be 
\begin{equation}
|\sigma\rangle  =  |\mathbf{q}\rangle + \sum_{i} \int \frac{d^3 \mathbf{q}_1}{(2\pi)^3} \frac{\exp(-i\mathbf{\mathbf{q}}_1\cdot \mathbf{X}_i) U(|\mathbf{q}-\mathbf{q}_1|)}{ E(\mathbf{q}) - E(\mathbf{q}_1)} |\mathbf{q}_1\rangle \mathrm{,}
\label{eq:eigen2}
\end{equation} 
where $U(\mathbf{q})$ is the Fourier transform of the ion-electron potential.
In the absence of the electron screening, $U(\mathbf{q}) = 4 \pi Ze^2/|\mathbf{q}|^2$, and
Eqs.~(\ref{eq:adler}) and (\ref{eq:eigen2}) give the susceptibility
\begin{equation}
 \alpha(\mathbf{k},\omega) = \alpha_{\mathrm{rpa}}(\mathbf{k},\omega) + \alpha_{\mathrm{dense}}(\mathbf{k},\omega) \mathrm{,} \nonumber 
\end{equation} 
where $\alpha_{\mathrm{rpa}}$, the well-known Lindhard susceptibility, is
\begin{equation}
\alpha_{\mathrm{rpa}} = \int \frac{d^3 \mathbf{q}}{ (2\pi)^3 } \frac{f(\mathbf{k}+\mathbf{q}) - f(\mathbf{q}) }{ \hbar \omega - E(\mathbf{k}+\mathbf{q}) + E(\mathbf{k}) } \mathrm{,} \label{eq:lind}
\end{equation}
and the $\alpha_{\mathrm{dense}}$, the dense plasma correction, is
\begin{eqnarray}
\alpha_{\mathrm{dense}}(\mathbf{k}, \omega) = \int \frac{d^3 \mathbf{s}}{(2\pi)^3} n_I(\mathbf{s})  \int \frac{d^3 \mathbf{q}}{ (2\pi)^3} \frac{U^2(\mathbf{s})}{A^2(\mathbf{q},\mathbf{k},\mathbf{s})} \nonumber \\
\times 
 \frac{ f(|\mathbf{q}+\mathbf{k}+\mathbf{s}|) - f(\mathbf{q})}{ \hbar \omega  - E(|\mathbf{q}+\mathbf{k}+\mathbf{s}|) + E(\mathbf{q})} \mathrm{,} \nonumber \\ \label{eq:eigen3}
\end{eqnarray}
where $A(\mathbf{q},\mathbf{k},\mathbf{s})$ is
\begin{eqnarray}
A^{-1}(\mathbf{q},\mathbf{k},\mathbf{s}) = \frac{1}{E(\mathbf{q}) - E(\mathbf{q}+\mathbf{s})}\nonumber \\
 - \frac{1}{E(\mathbf{q}+\mathbf{k}) - E(\mathbf{q}+\mathbf{k}+\mathbf{s})} \mathrm{.}  \nonumber \\ \label{eq:a}
\end{eqnarray}
The imaginary part of $\alpha_{\mathrm{dense}}$ is given as
\begin{eqnarray}
\mathrm{Im}\left[ \alpha_{\mathrm{dense}}\right] = 
\int \frac{d^3 \mathbf{s}}{(2\pi)^3} n_I(\mathbf{s})  \int \frac{d^3 \mathbf{q}}{ (2\pi)^3}\frac{U^2(\mathbf{s})}{A^2(\mathbf{q},\mathbf{k},\mathbf{s})}  \nonumber \\
\times  \pi \delta ( \hbar \omega  - E(\mathbf{q}+\mathbf{k}+\mathbf{s}) + E(\mathbf{q})) \nonumber \\ 
\times \left(f(\mathbf{q}+\mathbf{k}+\mathbf{s}) - f(\mathbf{q})\right)\mathrm{.} \nonumber \\
\label{eq:eigen4}
\end{eqnarray}
Some comments are made on Eq.~(\ref{eq:eigen3}), which is the main result of this paper. 
First, the integration over $\mathbf{s}$ corresponds to the summation of the
contribution from the Umklapp process in metals.
An electron with an initial momentum $\hbar \mathbf{k}$ absorbs the energy
$\hbar \omega$ and the momentum $\hbar(\mathbf{k}+\mathbf{s})$.  
This absorption is accompanied by an exchange of the momenta between an electron and an ion. 
Second, in metals, the integration over $\mathbf{s}$ is
replaced by the summation over the reciprocal lattice vectors, which represents
the summation of the inter-band transitions in the case of metals.
Third, the ion correlation is incorporated into $n_I(\mathbf{s})$.
The ions tend to be strongly correlated in the warm dense matter, which could be
easily accounted for by this formalism.
Fourth, from Eq.~(\ref{eq:die}), the decay of the Langmuir wave is given as
\begin{equation}
\frac{\gamma}{\omega} = \frac{1}{2}\mathrm{Im} \left[ \epsilon\right] = \frac{1}{2}\frac{4 \pi e^2}{ k^2} \mathrm{Im} \left( 
 \alpha_{\mathrm{rpa}} + \alpha_{\mathrm{dense}} \right) \mathrm{.}\label{eq:die2}
\end{equation}
The first term in the right hand side accounts for the Landau damping, and the second
term is the correction due to the presence of the ions.
Lastly, $\alpha_{\mathrm{dense}}$ does not arise purely from the quantum-mechanical effects,
as this term does not vanish in the limit $\hbar =0$. 
In the limit $\hbar = 0 $ and $k=0$, the decay rate becomes proportional to $n_I$
and $T_e^{-3/2}$, which is reminiscent of the electron-ion collision frequency.
However, the exact coefficient, which is derived in Eq.~(\ref{eq:eigen4}),
cannot be obtained from the ion-electron collision argument.
When $\hbar \neq 0$, the quantum degeneracy and the electron diffraction effects
are self-consistently incorporated into Eq.~(\ref{eq:eigen4}).  
  
\section{Degenerate Electron Case}
Consider the dense plasma where the electrons are completely degenerate.
The Fermi-energy is given as $ E_F =  36.4 \times (n/n_{24})^{2/3} \mathrm{eV}$,
where  $n_{24} = 10^{24} \mathrm{cm^{-3}}$.
The computation in this Section is valid as long as $T_e < E_F$.
In the partially degenerate and classical case, the electron screening and the quantum
diffraction need to be treated carefully, which would be considered elsewhere~\cite{sonprep}.  

\begin{figure}
\scalebox{0.4}{
\includegraphics{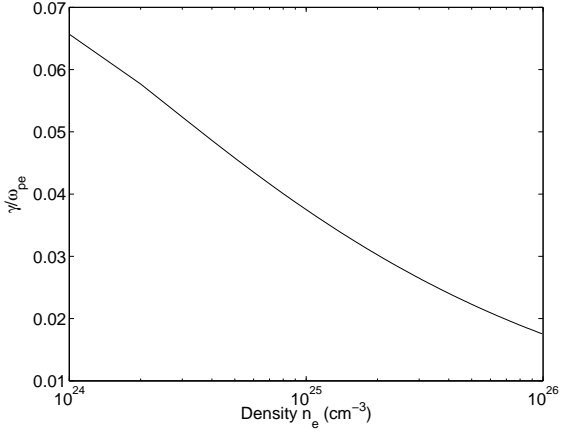}}
\caption{\label{fig:2} The damping rate of a degenerate hydrogen plasma for
a range of the electron density, $10^{24} \mathrm{cm^{-3}} < n_e < 10^{26} \mathrm{cm^{-3}} $. 
}
\end{figure}

The susceptibility $\alpha_{\mathrm{rpa}}$ of the free electron plasma, computed by
Lindhard~\cite{Lindhard}, is given as 
\begin{equation}
\alpha_{\mathrm{rpa}}(\mathbf{k},\omega) = \frac{3 n_e}{m_e v_F^2} h(z, u) \mathrm{,} \nonumber 
\end{equation}    
where $v_F = \sqrt{2E_F/m_e}$ is the Fermi velocity, $ z = k / 2 k_F$, $u = \omega /k v_F$, and $h= h_r + i h_i$. The real part of $h$, $h_r$, is
\begin{eqnarray}
h_r = \frac{1}{2} + \frac{1}{8z}\left( 1- (z-u)^2\right) 
\log \left( \frac{|z-u+1|}{|z-u+1|} \right) \nonumber  \\
+ \frac{1}{8z}\left( 1- (z+u)^2\right) 
\log \left( \frac{|z+u+1|}{|z+u+1|} \right),\nonumber \\ \label{eq:re} \nonumber 
\end{eqnarray}
and the imaginary part $h_i$ is
\begin{eqnarray}
h_i &=&  \frac{\pi}{2} u \mathrm{,} ~~~~ \mathrm{for~~}  |z+u| < 1   \mathrm{,} \nonumber \\
&=&  \frac{\pi}{8z} (1 - (z-u)^2) \mathrm{,} ~~~ \mathrm{for~~} |z-u| < 1 <  |z+u|  \mathrm{,}\nonumber \\ 
&=& 0   \mathrm{,} ~~~~~~~ \mathrm{if~~}  |z-u| > 1 \mathrm{,} \nonumber \\ \nonumber  \label{eq:im}
 \end{eqnarray}
In the limit $k = 0$, due to the delta function in Eq.~(\ref{eq:eigen4}),
$A^{-1}(\mathbf{q},\mathbf{k},\mathbf{s})$ in Eq.~(\ref{eq:a}) can be approximated as 
\begin{equation}
  A^{-1}(\mathbf{q}, \mathbf{k}, \mathbf{s}) = \frac{ \mathbf{k} \cdot \mathbf{s}}{m_e  \omega^2} \mathrm{,}  \nonumber 
\end{equation}
using the screening potential $U(s) = 4 \pi Z e^2 / (s^2 + k_{tf}^2) $,
where $k_{tf} = \sqrt{3} \omega_{\mathrm{pe}}/v_F $ is the Thomas-Fermi screening length.
In this case, $A$ is independent of $\mathbf{q}$, which can be taken out of
the $\mathbf{q}$-integration in Eq.~(\ref{eq:eigen3}). 
Using Eq.~(\ref{eq:lind}), Eq.~(\ref{eq:eigen4}) reads 
\begin{eqnarray}
 \mathrm{Im} \left[\alpha_{\mathrm{dense}}(\mathbf{k},\omega)\right] =   \int \frac{d^3 \mathbf{s}}{ (2\pi)^3 } \frac{\hbar^4(\mathbf{k} \cdot \mathbf{s})^2 (4 \pi Ze^2)^2 }{m_e^2 (\hbar\omega)^4 (s^2 + k_{tf}^2)^2 } \nonumber \\ \nonumber \\
\times n_I(\mathbf{s}) \mathrm{Im} \left[ \alpha_{\mathrm{rpa}}(\mathbf{k} + \mathbf{s}, \omega) \right] \mathrm{,}\nonumber \\ \label{eq:zero2} \nonumber 
\end{eqnarray}
which can be further simplified to 
\begin{eqnarray}
 \mathrm{Im} \left[\alpha_{\mathrm{dense}}(\mathbf{k}, \omega)\right]
= \frac{4 E_F^2 Z (k_F e^2)^2 }{(\hbar\omega)^4 }\frac{(4\pi)^2}{ 3 \pi^2 } \int \frac{d^3 \mathbf{s}}{ k_F^3(2\pi)^3 }   \nonumber \\ \nonumber \\
\times \frac{(\mathbf{k}\cdot \mathbf{s})^2}{ k^2 (s^2 + k_{tf}^2)^2 } \mathrm{Im} \left[\alpha_{\mathrm{rpa}} (\mathbf{k}+\mathbf{s}, \omega) \nonumber\right] \mathrm{,} \\ \label{eq:kzero}
\end{eqnarray} 
where we assumed $n_I(\mathbf{s}) = n_I$. 
$\eta(0) $ in Eq.~(\ref{eq:lang}) from the above equation is shown in Fig.~\ref{fig:2}.

\begin{figure}
\scalebox{0.4}{
\includegraphics{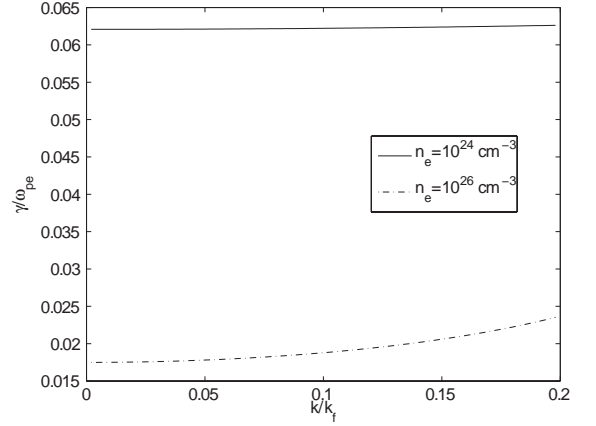}}
\caption{\label{fig:3}
The damping rates of a degenerate hydrogen plasma as a function of the wave vector
for two cases of electron density.
}
\end{figure}

For non-zero $k$'s, Eq.~(\ref{eq:kzero}) cannot be used, as $A$ is dependent
on $\mathbf{q}$. A rather complicated expression for $A$ is
\begin{eqnarray}
&A^{-1}&(\mathbf{q}, \mathbf{k}, \mathbf{s} )=
\frac{\hbar^2 \mathbf{k} \cdot \mathbf{s}}{m_e(\hbar \omega)^2} \nonumber \\ 
&\times& \left[
\left( 1 - \frac{\hbar \mathbf{k} \cdot (\mathbf{q} + \mathbf{s}/2)}{m \omega}\right) 
\left( 1 - \frac{\hbar \mathbf{k} \cdot (\mathbf{q} - \mathbf{s}/2)}{m \omega}\right) 
\right]^{-1} \mathrm{,} \nonumber 
\end{eqnarray} 
which should remain in the integrand for the $\mathbf{q}$-integration in
Eq.~(\ref{eq:eigen3}).
For given $\mathbf{s}$ and $\mathbf{k}$, we integrate over $\mathbf{q}$
first and then over $\mathbf{s}$.
For simplicity, we assume $\mathbf{k} = (k_x, 0, 0)$ and $\mathbf{s} = (s\cos\theta, s \sin\theta, 0)$, and denote $\mathbf{q} = q_1 \hat{x}_1
+ q_2 \hat{x}_2 + q_3 \hat{x}_3$, where $\hat{x}_1 = (k+ s \cos\theta, s\sin\theta, 0)/|\mathbf{k}+\mathbf{s}|$, $ \hat{x}_2 =  (- s\sin\theta, 1 + \cos\theta, 0)/|\mathbf{k}+\mathbf{s}|$,  and $ \hat{x}_3 = (0,0, 1)$.
In this coordinate system, $ A $ depends only on $k$, $s$, $\theta$, and $q_2$: 
\begin{equation} 
A^{-1}(\mathbf{k},\mathbf{s}, q_2) = \frac{\hbar^2 \mathbf{k} \cdot \mathbf{s}}{m_e(\hbar \omega)^2}  \kappa^{-} \kappa^{+} \label{eq:sp} \nonumber \mathrm{,} 
\end{equation}
where 
\begin{equation}
\frac{1}{\kappa^{\pm}}  = 1 - \frac{k^2 + sk \cos\theta}{|\mathbf{k}+\mathbf{s}|^2}
+ \frac{\hbar}{m_e \omega}\left( 
\frac{sk\sin\theta q_2}{|\mathbf{k}+\mathbf{s}|} \pm  \frac{sk \cos\theta}{2}\right) \nonumber  \mathrm{.}
\end{equation}
The real part of $\alpha$ remains still complicated, but the imaginary part given
in Eq.~(\ref{eq:eigen4}) could be simplified as the delta function
would eliminate $q_1$ integration and the integrand of $dq$ is independent of $q_3$.
We do not present the detailed steps here.
After some tedious manipulation, the $\mathbf{q}$-integration is reduced to an one-dimensional
integration
\begin{eqnarray}
\int \frac{d^3 \mathbf{q}}{ (2\pi)^3}\frac{U^2(\mathbf{s})}{A^2(\mathbf{q},\mathbf{k},\mathbf{s})} \mathrm{Im} \left[ \frac{f(\mathbf{q}+\mathbf{k}+\mathbf{s}) - f(\mathbf{q})}{  \hbar \omega  - E(\mathbf{q}+\mathbf{k}+\mathbf{s}) + E(\mathbf{q})}\right] \nonumber \\  \nonumber \\ \nonumber \\
= \frac{(\hbar^2 \mathbf{k} \cdot \mathbf{s})^2 U^2(\mathbf{s})}{m_e^2(\hbar \omega)^4}  \int_{-\sqrt{1-(z-u)^2}}^{\sqrt{1-(z-u)^2}} dq_2 
  \frac{m_e}{4\pi^2 \hbar^2 |\mathbf{s}+\mathbf{k}|}  \nonumber 
 \\ \nonumber \\ \times \kappa^{+} \kappa^{-1} \left(R^-(k_F, z, u, q_2)-R^+(k_F, z,u, q_2)\right) \mathrm{,} \nonumber \\  \label{eq:comp} \nonumber 
\end{eqnarray}
where $ R^{\pm}(k_F,z,u,q_2)  =  \sqrt{ k_F^2(1 - (z\pm u)^2 ) - q_2^2 }$  if $ k_F^2(1 - (z\pm u)^2 > 0 $,  $R^{\pm}(k_F,z,u,q_2) = 0$ ~ if  $k_F^2(1 - (z\pm u)^2 \le 0 $, $z = |\mathbf{k}+\mathbf{s}|/ 2k_F$,
and  $u = \omega / kv_F$. The right-hand side of the above equation is a function only
of $k$, $s$, and $\theta$.
We do the $\mathbf{s}$-integration in the spherical coordinate and arrive at
\begin{eqnarray}
 \mathrm{Im} \left[\alpha(\mathbf{k}, \omega)\right] = \int \frac{2\pi s^2 d\mu ds}{(2\pi)^3 } n_I(\mathbf{s}) \frac{(\hbar^2 ks \mu)^2 U^2(\mathbf{s})}{m_e^2(\hbar \omega)^4} \nonumber \\ \nonumber \\
\times  \int_{-k_F \sqrt{1-(z-u)^2}}^{k_F \sqrt{1-(z-u)^2}} dq_2   \frac{m_e}{4\pi^2 \hbar^2 |\mathbf{s}+\mathbf{k}|} \nonumber \kappa^{+} \kappa^{-1} \nonumber  \\  \nonumber \\ \times 
   \left(R^-(k_F, z, u, q_2)-R^+(k_F, z,u, q_2)\right)  \mathrm{.}\nonumber \\  \label{eq:comp2}  \nonumber 
\end{eqnarray}
This is a three-dimensional integration that needs to be evaluated numerically.
In Fig.~(\ref{fig:3}), $\eta $ in Eq.~(\ref{eq:lang}) as a function of $k$ is shown for a hydrogen plasma
of two different electron densities.

\section{Raman Scattering}

In this Section, we discuss the existence of a regime where an x-ray pulse
can be compressed by the BRS.
More detailed presentation can be found in Ref.~\cite{sonpre} for the case of metals.
Here we provide a similar computation in the case of degenerate dense hydrogen plasmas.

The BRS is a three-wave interaction where two light
pulses and one Langmuir wave exchange the energy~\cite{Murray,McKinstrie,McKinstrie2}.
A one-dimensional interaction can be written as~\cite{McKinstrie,McKinstrie2}:
\begin{eqnarray}
\left( \frac{\partial }{\partial t} + v_1 \frac{\partial}{\partial x} + \nu_1\right)A_1  = -ic_1 A_2 A_3  \nonumber \mathrm{,}\\
\left( \frac{\partial }{\partial t} + v_2 \frac{\partial}{\partial x} + \nu_2\right)A_2  = -ic_2 A_1 A^*_3   \label{eq:2} \mathrm{,} \\
\left( \frac{\partial }{\partial t} + v_3 \frac{\partial}{\partial x} + \nu_3\right)A_3  = -ic_3 A_1 A^*_2  
\nonumber \mathrm{,}
\end{eqnarray}
where $A_i= eE_i/m_e\omega_ic$ is the ratio of  the electron quiver velocity of the pump pulse ($i=1$)
and seed pulse ($i=2$), relative to the velocity of the light $c$, and $A_3 = \tilde{n_e}/n_e$ is
the intensity of the Langmuir wave.
$\nu_1 $ ($\nu_2$) is the rate of the inverse bremsstrahlung of the pump (seed), $\nu_3$ is
the plasmon decay rate, $ c_i = \omega_{pe}^2/2\omega_i$ for $i=1$ and 2, $c_3 = (cq)^2/2\omega_3$,
and $\omega_{1} (\omega_{2})$  is the frequency of the pump (seed).
In the BRS (FRS), the energy conservation is given as $\omega_1 = \omega_2 + \omega_{pe}$
($\omega_1 = \omega_2 + \omega_{pe}$) and the momentum conservation is given as
$ q = k_1 -k_2 $ ($ q = k_1 +k_2 $).

The inverse bremsstrahlung ($\nu_1$ and $\nu_2$) in metals is given as \cite{shima, son2},
\begin{equation}
\nu(\omega, E) = 4 \pi n_I Z^2  \frac{e^4}{m_e^2 v_F^3} \frac{\omega_{pe}^2}{\omega^2} \frac{F(\alpha)}{s^2 \kappa^{1/2}}\mathrm{,}
\label{eq:brm}
\end{equation}  
where $ \kappa =  (\hbar \omega / 2 m_e v_F^2) < 1 $, $v_F$ is the Fermi energy and $ \alpha^2 = 2e^2E^2/m_3\hbar \omega^3 $ with $E$ being the electric field of the pulse, and 
the function $ F(\alpha) $ can be approximated as $\alpha^2/6 $ when $\alpha < 2$ and $(2/\pi\alpha)\log(2\alpha)^2$ otherwise~\cite{shima}.
In Eq.~(\ref{eq:brm}), it is assumed that $\hbar \omega > E_F$ and $\hbar \omega > T_e $ where $T_e $ ($E_F$) is the electron temperature (Fermi energy).  
The above equation can be simplified, when $\alpha < 2$,  as 
\begin{equation}
 \nu = 2.9 \times 10^{15} Z \kappa^{-3/2} \frac{\omega_{\mathrm{pe}}^2}{\omega^2}/ \label{eq:brem} \sec \mathrm{.}
\end{equation}
If $1/\nu $ in Eq.~(\ref{eq:brem}) is shorter  than the pulse duration, the pump pulse will
be heavily damped before being compressed.
Note that  $\nu_1$ is proportional to $\omega_{\mathrm{pe}}^2/\omega_1^2$.

From Eqs.  (\ref{eq:lang}) and (\ref{eq:2}), the BRS growth rate, assuming that the pump intensity is large enough to make the BRS unstable, is roughly estimated to be  
\begin{equation}
g_B =  \frac{A_1^2}{\eta(q)} \frac{\omega_2 }{\omega_{\mathrm{pe}}} \label{eq:g} \omega_{\mathrm{pe}} / \sec \mathrm{.}
\end{equation} 
The larger $g_B$ is,  the stronger the BRS is.   

Now, a possible operating regime for the Raman compression is obtained from
Eqs.~(\ref{eq:lang}), (\ref{eq:2}),  and (\ref{eq:brem}).  
The stability condition for the FRS is given, from Eq.(\ref{eq:2}),  as 
$ c_2c_3|A_1|^2 <  \nu_2 \nu_3 $, where $\nu_2$ is obtained from Eq.~(\ref{eq:brem})
and $\nu_3$ is from Eq.~(\ref{eq:lang}), and $c_3= \omega_{\mathrm{pe}}/2$.
This condition can be written as $|A_1| > |A_{1F}|$ where 
\begin{equation}
 |A_{1F}|^2 = \frac{4 \nu_2(q_F) \nu_3}{\omega_{\mathrm{pe}}^2} \frac{\omega_2}{\omega_{\mathrm{pe}}}
 \label{eq:for} \mathrm{,}
\end{equation} 
and $q_F = \omega_{\mathrm{pe}}/c$.  
The same condition for the BRS can be estimated, using
$c_3 =  (\omega_1 +\omega_2)^2/2\omega_{\mathrm{pe}}$, as $|A_1| < |A_{1B}|$, where 

\begin{equation}
 |A_{1B}|^2 = \frac{\nu_2(q_B) \nu_3}{\omega_{\mathrm{pe}}^2} \frac{4\omega_{\mathrm{pe}} \omega_2}{(\omega_2 +\omega_1)^2}
 \label{eq:back} \mathrm{.}
\end{equation} 
and $q_B = (\omega_1 +\omega_2)/c$.  
$|A_{1B}|^2$ in Eq.~(\ref{eq:back}) is smaller than $ |A_{1F}|^2$ in Eq.~(\ref{eq:for})
by a factor of $(\omega_1 +\omega_2)^2/\omega_{\mathrm{pe}}^2 (\eta(q_F)/\eta(q_B))$,
which is very large since  $\eta(q_F)/\eta(q_B) \cong 1$. 
Therefore there exists a regime where the BRS is unstable but the FRS is  stable.  
This is due to the strong inverse bremsstrahlung and  a rather slowly varying
damping of the Langmuir waves as a function of $k$.

We provide an estimation for  hydrogen plasmas with $n_e = 10^{24} \mathrm{cm^{-3}} $
and $10^{26} \mathrm{cm^{-3}} $. 
In the former case, when $ \hbar \omega_{\mathrm{pe}} = 37.11 \ \mathrm{eV} $ and
$\omega_1 / \omega_{\mathrm{pe}}  = 10 $, it is estimated that $\eta(k) \cong 0.06 $
from Fig.~\ref{fig:3}.
The threshold for the FRS estimated from Eq.~(\ref{eq:for}) is $ |A_{1F}|^2 = 3 \times 10^{-4}$,
which corresponds to $3.4  \times 10^{19} \ \mathrm{W}/\mathrm{cm}^2 $.
The threshold for the BRS estimated from Eq.~(\ref{eq:back}) is
$|A_{1B}|^2 = 0.0075 \times 10^{-4}$ or $ 8.4 \times 10^{16} \  \mathrm{W} /\mathrm{cm}^2 $. 
For $ n_e = 10^{26} \mathrm{cm^{-3}} $, $ \hbar \omega_{\mathrm{pe}} = 371.19 \ \mathrm{eV} $, and
$\omega_1 / \omega_{\mathrm{pe}}  = 10 $, $\eta(0) \cong 0.18 $  and $\eta(k= 2 \omega/c) \cong 0.25 $
from Fig.~\ref{fig:3}.
The threshold for the FRS estimated from Eq.~(\ref{eq:for}) is $ |A_{1F}|^2 =0.288 \times 10^{-4}$
which corresponds to $3.225 \times 10^{20} \  \mathrm{W}/\mathrm{cm}^2 $.
The threshold   for the BRS estimated from Eq.~(\ref{eq:back}) is
$|A_{1B}|^2 = 0.000998 \times 10^{-4}$ or $ 1.116 \times 10^{18} \ \mathrm{W} /\mathrm{cm}^2 $.

\section{Conclusion}

The prevalent Landau damping theory predicts the plasmon decay rate to vanish
in the long wavelength limit.
On the other hand, the observations from the electron stopping experiments in metals
suggest that it is finite in that limit.
Using a well-known dielectric function formalism developed in the condense matter,
we propose a theoretical framework predicting the plasmon decay rate in dense plasmas
more accurately (Eq.~(\ref{eq:eigen3})) than conventional theories.
A few advantages of our approach are in order:
First, the robustness of this framework is verified through the experiments with metals~\cite{Sturm}.
Second, the quantum degeneracy and the diffraction of the electrons are self-consistently
accounted for.
Third, the inter-ion correlations could be incorporated for a wide range of
physical parameter regimes through $n_I(\mathbf{s})$, the static two-point correlation
function among the ions.
Lastly, the theory  takes into account of the Umklapp process.  
The computation based on this dielectric function approach suggests that the plasmon decay
rate in dense plasmas in the long wavelength limit is much higher than the prediction by
other theories.

This strong decay has implications for various processes in dense plasmas,
warm dense matter and metals.   
For instance, it renders the FRS less harmful for the Raman compressor
in metals and dense plasmas, compared to the FRS in rare dense plasmas.
The plasmon decay in metals strongly depends on the direction relative to the lattice structure. 
As the electron temperature $T_e$ increases, the electron screening of the ions
gets weaker, resulting in a stronger plasmon damping~\cite{Sturm,Sturm2}.
As $T_e$ increases further, the ions would finally lose their lattice structure.
A larger value of $\eta(k)$ might be preferred for a strong pump, as it would weaken
the FRS; however, a small value of $\eta(k)$ might be preferred for a weak pump,
as a strong BRS is necessary. 

As a consequence, there exist a parameter regime in metals where
the BRS compression is strong while the FRS is stable (Sec. VI).
As an example, a pump pulse with a duration of hundreds of femto seconds could be compressed,
via a plasmon, to a pulse of a few or sub-femto second without FRS pump depletion~\cite{sonpre}.   
The computation of the decay rate for a wide range of parameter regimes, by extending
our formalism given in Eq.~(\ref{eq:eigen4}), may lead to detailed engineering
of the plasmon decay for the x-ray BRS compression.





The plasmon decay in dense plasmas is a complicated process.
The Umklapp process dominates for low $k$ as presented here, and the Landau damping
dominates for high $k$.
In addition, the degeneracy and diffraction would suppress the damping~\cite{sonlandau}.   
Experiments measuring the damping rate in warm dense matter might be readily
available by observing the loss of the electron energy in the thin heated foil
experiment~\cite{Gibbons}. 
The effect of the phase transition on the plasmon decay is theoretically challenging,
but is important for its practical application in the BRS x-ray compression and
other processes.

\bibliography{plasmon}

\end{document}